\documentstyle[aps,prb,epsfig,floats]{revtex}
\begin{document}

\twocolumn[\hsize\textwidth\columnwidth\hsize\csname
@twocolumnfalse\endcsname

\title{A first-principles study of the structure and lattice dielectric response
of $\mbox{CaCu}_{3}\mbox{Ti}_{4}\mbox{O}_{12}$}
\author{Lixin He, J. B. Neaton, Morrel H. Cohen, and David Vanderbilt}
\address{Department of Physics and Astronomy, Rutgers University,\\
Piscataway, New Jersey 08855-0849}
\author{C. C. Homes}
\address{Department of Physics, Brookhaven National Laboratory,\\
Upton, NY 11973-5000}
\date{January 18, 2002}

\draft
\maketitle

\begin{abstract}
Structural and electronic properties of CaCu$_3$Ti$_4$O$_{12}$
have been calculated using density-functional theory within the
local spin-density approximation.  After an analysis of structural
stability, zone-center optical phonon frequencies are evaluated
using the frozen-phonon method, and mode effective charges are
determined from computed Berry-phase polarizations. Excellent
agreement between calculated and measured phonon frequencies
is obtained; calculated mode effective charges are in poorer
agreement with experiment, although they are of the correct
order of magnitude; and the lattice contribution to the static
dielectric constant is calculated to be $\sim$40. On the basis
of these results, various mechanisms are considered for the
enormous dielectric response reported in recent experiments.
No direct evidence is found for intrinsic lattice or electronic
mechanisms, suggesting that increased attention should be given to
extrinsic effects.
\end{abstract}

\pacs{PACS: 77.84.-s, 63.20.-e, 71.20.-b, 77.22.-d}

\narrowtext

]

\section{INTRODUCTION}

Recently CaCu$_3$Ti$_4$O$_{12}$ (CCTO) was discovered to possess
one of the largest static dielectric constants ever measured,
reaching nearly $\epsilon_0\sim80,000$ for single-crystal samples at room
temperature.\cite{sub,ram,homes}  Remarkably, the enormous static
susceptibility is almost {\it constant} over a wide temperature
range ($\sim$0-500\,K).
Moreover, measurements of the dynamic
susceptibility\cite{homes} show that $\epsilon$ falls from its huge
static value to a more conventional $\epsilon\sim100$ with increasing
frequency, and that this crossover shows a Debye-like frequency
dependence (e.g., $\tau^{-1}\simeq1$\,kHz at 80\,K) with a relaxation
rate $\tau^{-1}$ exhibiting an activated dependence on temperature.
However, the physical origins of the enormous $\epsilon_0$ and
the related Debye relaxation process remain mysterious.

It is tempting to classify a solid exhibiting such a prodigious
dielectric response as a relaxor or a ferroelectric, but mounting
empirical evidence tends to exclude CCTO from either category.
The low-temperature static dielectric response of perovskite
ferroelectrics can surpass 1000,\cite{lg} rising rapidly above
this value with increasing temperature and peaking
near the ferroelectric phase transition temperature.  However, this peak in
response is generally confined to a limited temperature range
around the transition, whereas that seen in CCTO appears quite
temperature independent (at fixed low frequency) over a much
wider range.  Furthermore, high-resolution x-ray and neutron
powder diffraction measurements on CCTO reveal a centrosymmetric
crystal structure (space group $Im3$) persisting down to 35\,K,
ruling out a conventional ferroelectric phase transition.
The Debye-Waller factors are also found to be normal over a wide
temperature range, from 35\,K to beyond room temperature,\cite{sub}
inconsistent with random local polar displacements of the magnitude
typical in ferroelectrics or their disordered paraelectric phases.
Nanodomains or disorder effects, common to relaxor materials, are
notably absent: neither superstructure peaks nor strong diffuse
scattering are observed in diffraction experiments. Single-crystal
samples were described to be mainly twinned, i.e., containing
nanoscale domains \cite{sub} differing in the sign of the rotation
of the TiO$_6$ octahedra. The presence of the associated domain
boundaries may have important consequences for the dielectric
properties, as we shall discuss further below.

\begin{figure}
\begin{center}
\epsfig{file=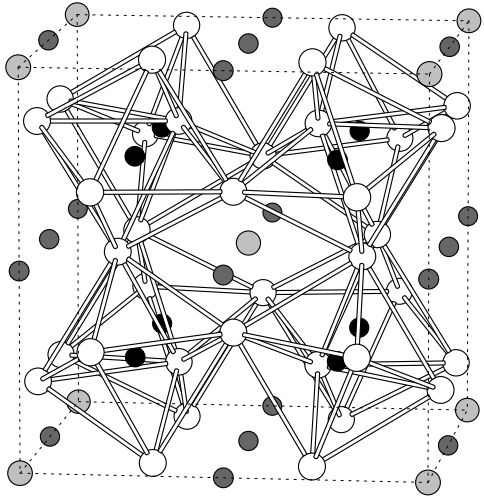,width=2.5in}
\end{center}
\caption{Structure of CaCu$_3$Ti$_4$O$_{12}$ (CCTO), showing tilted
oxygen octahedra.  White, light, dark, and black atoms are
O, Ca, Cu, and Ti respectively.  Dashed lines indicate 40-atom primitive
cell of antiferromagnetic spin structure.}
\label{fig:structure}
\end{figure}

The crystal structure of CCTO is body-centered cubic with four
ATiO$_3$ perovskite-type formula units per primitive cell
(where A is either Ca or Cu).  The doubled conventional
40-atom unit cell is shown in Fig.~\ref{fig:structure}.
The local moments associated with $d$ holes on the Cu$^{2+}$
cations result in the formation of long-range magnetic
order; transport, Raman, and neutron diffraction experiments
indicate that CCTO is an {\it antiferromagnetic insulator}
having a N\'eel temperature $T_N$ of $25$\,K (and Weiss constant
$\theta_w=-34$\,K).\cite{sub,ram,ram2,ram3}  Thus, the 40-atom
simple-cubic cell in Fig.~\ref{fig:structure} is also the primitive
cell of the spin structure.
Although the extraordinary dielectric behavior persists far above
the temperature scale of the magnetic order, the influence of
the strongly-correlated local moments, which will likely remain
above $T_N$, is currently unknown.

The purpose of this study is to examine rigorously the ground-state
properties of this material using first-principles calculations
in order to identify or exclude possible mechanisms underlying
the large dielectric response.  Our calculations within the
local spin-density approximation can serve as an important first step
toward the eventual elucidation of this unusual phenomenon. After
briefly discussing the technical aspects of our first-principles
calculations in Sec.~II, we provide a detailed analysis of the
ground-state structural and electronic properties in Sec.~III. In
Sec.~IV we present the results of frozen-phonon calculations
from which we determine the frequencies of the zone-center
optical modes. As emphasized below, all modes are found to be
stable, consistent with the absence of a structural transition
at low temperature.  Our calculated frequencies agree well with
those measured by Homes {\it et al.},\cite{homes} although
we predict one extra mode apparently absent from their data.
The phonon contribution to the static dielectric constant we
calculate from first principles does not agree with empirical
observation nearly as well as our computed phonon frequencies,
but it nonetheless underestimates the measured
contribution from the IR-active lattice modes seen in recent
experiments ($\sim$80)\cite{homes} by a factor less than three.  In Sec.~V we discuss
possible explanations for the observed enormous static dielectric
response and associated Debye relaxation behavior, focusing
attention on possible extrinsic mechanisms as well as considering
intrinsic lattice and electronic mechanisms.  We conclude in Sec.~VI.

\section{METHODOLOGY}

To investigate the ground-state structural, electronic,
and lattice-dynamical properties of antiferromagnetic CCTO, we
employ density functional theory\cite{hk} within the local
spin-density approximation (LSDA)\cite{ks} via the Vienna {\it
ab-initio} Simulations Package (VASP);\cite{kresse1,kresse2}
VASP utilizes a plane-wave basis and Vanderbilt ultrasoft
pseudopotentials.\cite{vanderbilt} Our pseudopotentials include
non-linear core corrections \cite{froyen} and, for Ca and Ti,
these potentials treat the highest occupied $p$
shell electrons explicitly as valence; for Cu we also consider
electrons in the $3d$ and $4s$ shells self-consistently. A 37\,Ry
plane-wave cut-off results in convergence of the total energy to 1
meV/ion. The ions are steadily relaxed towards equilibrium until
the Hellmann-Feynman forces are less than 10$^{-2}$\,eV/{\AA}.
When calculating the phonon frequencies, we converge the forces to
$10^{-5}$\,eV/{\AA}, resulting in precision of less than
1~cm$^{-1}$. Brillouin-zone integrations are performed with a Gaussian
broadening of 0.1\,eV during all relaxations.  All calculations are
performed with a $2\times2\times2$ Monkhorst-Pack {\bf k}-point
mesh (equivalent to a $4\times4\times4$ mesh for a single 5-atom
perovskite unit cell); doubling the cut-off or the size of the
{\bf k}-point mesh does not significantly change the total energy
or forces.

\section{STRUCTURE}

The structure of CaCu$_3$Ti$_4$O$_{12}$ was first determined from
neutron powder-diffraction data over twenty years ago;\cite{Bochu}
these results have also been confirmed by more recent neutron-diffraction
measurements.\cite{sub} The primitive cell is
body-centered cubic, with space and point groups $Im3$ and $T_h$,
respectively, and contains 20 atoms; this space group includes
inversion symmetry, precluding the possibility of a net spontaneous
polarization or ferroelectricity in this structure.  Because there
are three $\mbox{Cu}^{2+}$ cations in the primitive 20-atom cell,
this structure cannot easily accommodate an antiferromagnetic (AFM)
spin structure.
In fact, neutron diffraction\cite{ram3} indicates that the
AFM state can be described with a doubled 40-atom simple-cubic
unit cell as shown in Fig.~\ref{fig:structure}.  In this structure,
each Cu--Cu nearest-neighbor pair has antiparallel spins.
All calculations are performed using this 40-atom cell, although
it should be kept in mind that the atomic coordinates and total
charge density have the periodicity of the 20-atom bcc cell.

\begin{table}
\caption{Comparison of calculated and measured structural parameters of CCTO. 
CCTO has space group $Im3$ (point group $T_h$); the Wyckoff positions are
Ca($0,0,0$), Cu($1/2,0,0$), Ti($1/4,1/4,1/4$), O($x,y,0$).
\label{tab:structure}}
\vskip 0.1cm
\begin{tabular}{ldd}
Structural parameter & LSDA       & Exp. (35\,K)\\
\hline
$a$ (a.u.)  &   7.290    & 7.384   \\
$x$         &   0.303    & 0.303   \\
$y$         &   0.175    & 0.179
\end{tabular}
\end{table}

Beginning with the experimental structural parameters, the 40-atom
cubic cell is first relaxed for a set of fixed volumes, and the
equilibrium lattice constant is then determined by minimizing
the total energy as a function of volume.  Our calculated
lattice constant is $a=7.290$\,\AA, less than the value of
{7.384\,\AA} measured at 35\,K by slightly more than 1\%, as
is typical in LDA calculations.  The internal parameters of the
structure are then relaxed at our calculated lattice constant;
their values appear in Table \ref{tab:structure}. The stability
of the $Im3$ structure is also checked by randomly displacing the
ions (lowering the point symmetry from $T_h$ to $C_1$) and allowing
them to relax steadily toward their equilibrium positions. We also
randomly displaced the ions {\it in the same Cartesian direction}, in
this case reducing the point symmetry to $C_{2v}$.  Importantly,
in both cases the ions returned to their high-symmetry ($Im3$) sites,
indicating the absence of unstable (i.e., imaginary-frequency)
zone-center optical phonons.  (Here we limit ourselves to lattice
distortions having the same random displacements in each simple cubic
unit cell. Thus both zone-center and (111) zone-boundary instabilities
have been investigated for the BCC structure.)

As is evident from Fig.~\ref{fig:structure}, the $Im3$ structure of
CCTO can be obtained from an ideal simple-cubic CaTiO$_3$ perovskite
by substituting 3/4 of the Ca ions by Cu in a bcc pattern and
rotating each TiO$_6$ octahedron by a fixed angle about one of
the four $\{111\}$ axes.  Each Ti ion is then fully coordinated with
six equidistant oxygen ions, calculated to be 1.94\,\AA~away.  Despite these rotations, the
Ca ion continues to possess 12 equidistant oxygen neighbors (calculated
distance 2.55\,\AA) as might be expected for A cations in
the ideal perovskite structure.  However, the tilted octahedra
do result in a radically different local environment for
each Cu$^{2+}$ ion: all copper ions are coordinated by a planar
arrangement of four nearest-neighbor oxygens (at 1.92\,\AA) and eight
oxygens that are considerably further away.  The tilting occurs to
relieve tension originating from the relatively small ionic radii
of both A-type cations. In this context we note that the ground
states of CaTiO$_3$,\cite{Cockayne} and to a lesser extent
SrTiO$_3$,\cite{Sai} also exhibit similar octahedral rotations.

\section{ELECTRONIC AND MAGNETIC STRUCTURE}

Our calculation of electronic structure, performed on the
fully-relaxed geometry, confirms that the ground state is an
antiferromagnetic (AFM) insulator (at $T$=0) and further illuminates
the bonding in CCTO.
We find the total charge density to be symmetric, and the spin
density to be antisymmetric, under the fractional lattice translation
$(a/2,a/2,a/2)$ consistent with the expected AFM state. (Here $a$ is
the lattice constant.)
\begin{figure}
\begin{center}
\epsfig{file=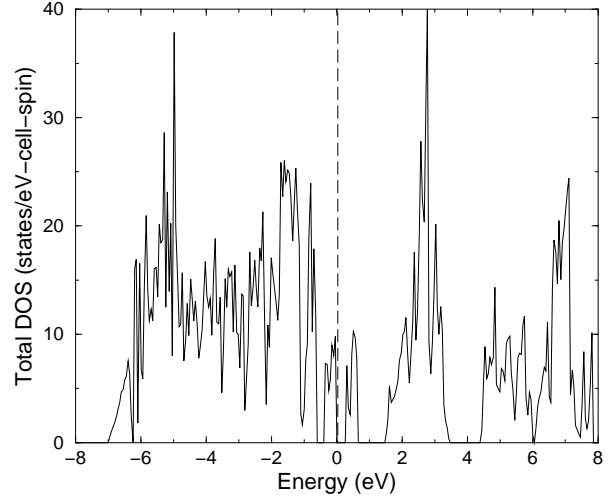,width=3.1in}
\end{center}
\caption{Density of states (DOS) for CCTO.
Spin-up and spin-down DOS are identical, as expected for an
antiferromagnet. Vertical dashed line indicates valence-band
maximum.}
\label{fig:totdos}
\end{figure}
We plot the density of single-particle states (DOS) in
Fig.~\ref{fig:totdos}, calculated within the LSDA for a single spin
channel. Both spin channels exhibit identical DOS, as expected for an
AFM insulator.  The bands in the energy range extending from $-$7
to 1\,eV are mainly O
$2p$ and Cu $3d$ in character.  The states deeper in these bands, from
$-$7 to$-$0.5\,eV, are found to consist mainly of Cu 3$d$ orbitals
that hybridize weakly with O $2p$ orbitals that point toward their Ti
neighbors.  In the range from $-$0.5\,eV to 1\,eV we find two narrow
bands, each containing just three weakly-dispersive states, composed
of strongly $\sigma$-antibonding combinations of Cu 3$d$ orbitals and 
O $2p$ orbitals pointing to the Cu from its four near oxygen neighbors
(more about this to follow below).  The unoccupied conduction-band states
above 1\,eV in Fig.~\ref{fig:totdos} are primarily Ti 3$d$-like, with
mainly $t_{2g}$ states making up the bands in the 1-4\,eV range and
mainly $e_g$ states above 4.5 eV and continuing up to about 7.5 eV.

\begin{figure}
\begin{center}
\epsfig{file=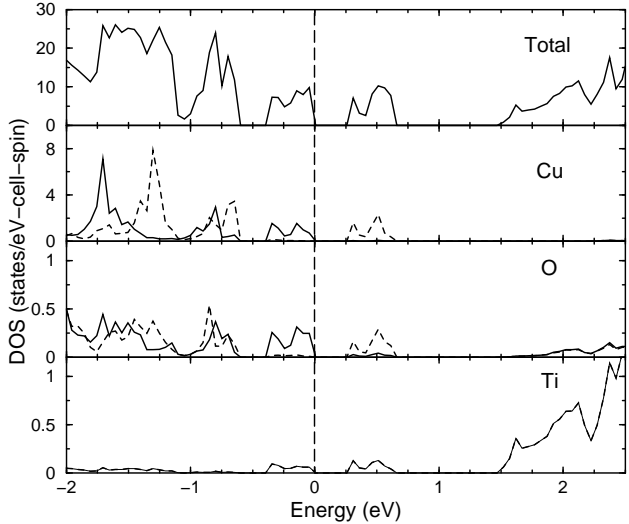,width=3.25in}
\end{center}
\caption{Total DOS (states/eV-cell-spin) and site-projected partial DOS
(PDOS) (states/eV-site-spin) for CCTO. Solid and dashed curves are
spin-up and spin-down densities, respectively.  Vertical dashed line
indicates valence-band maximum.  The PDOS is calculated using sphere
radii of 0.7\,\AA, 0.7\,\AA, and 1.25\,{\AA} for Cu, Ti, and O,
respectively.  The PDOS of Ca, not shown, is very small in this
energy range.}
\label{fig:PDOS}
\end{figure}

More details are provided in Fig.~\ref{fig:PDOS}, which affords a
magnified view of the total DOS near the fundamental gap and decomposes
the DOS into a site-projected partial DOS (PDOS) for the constituent Cu,
Ti, and O atoms in the crystal.  (The PDOS of Ca is quite small in this
energy range.)  By symmetry, the total DOS and the Ti PDOS are not
spin-polarized, but the spin splitting shows up clearly on Cu and O
sites where spin-up and spin-down densities are indicated by solid and
dashed curves respectively.  Figure \ref{fig:PDOS} clearly illustrates
that the fundamental band gap and local magnetic moment arise from the
splitting of the six weakly dispersive bands of Cu($3d$)--O($2p$)
antibonding orbitals in the $-$0.5-1\,eV energy range.

\begin{figure}
\begin{center}
\epsfig{file=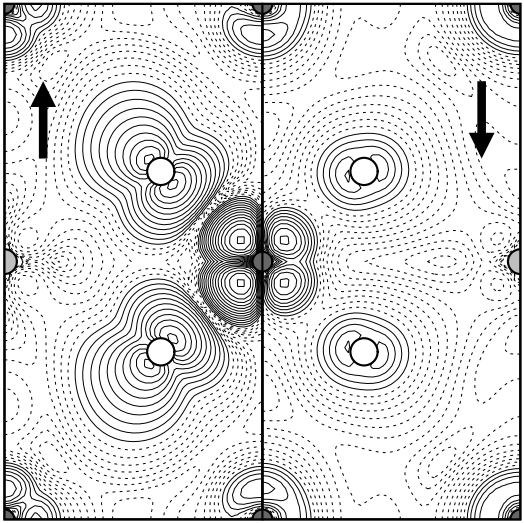,width=2.7in}
\end{center}
\caption{Spin-up (left panel) and spin-down (right panel)
charge densities associated with the three highest occupied valence
bands of CCTO plotted with logarithmic contours (five intervals per
decade) in the $z$=0 plane.  Each density actually has mirror symmetry
across the central dividing line.  To facilitate comparison of
magnitudes between panels, densities falling below an arbitrary
threshold are shown dashed.  Atoms are shaded as in Fig.~1.}
\label{fig:wavefunc}
\end{figure}

Further insight about these states can be gained by inspecting
Fig.~\ref{fig:wavefunc}, where a contour plot is shown of the spin-up and spin-down
charge densities associated with the lower three occupied valence-band
states in this group ($-$0.5-0\,eV); a plot for the three unoccupied
conduction-band states (0.2-0.7\,eV) would look similar but with
spins reversed.  Figure \ref{fig:wavefunc} clearly reveals the
Cu($3d$)--O($2p$) $\sigma$-antibonding nature of the states, which
extend over the central cluster composed of a Cu ion and its four close
oxygen neighbors at 1.92\,{\AA} (see Sec.~III).  This cluster evidently
forms a strongly spin-polarized unit with a ferromagnetic alignment
of the magnetic moments on the Cu and O atoms comprising it, indicating
a significant role of the oxygen atoms in the magnetic structure.
From the spin densities we estimate
the magnetic moment of each CuO$_4$ complex to be $\sim$0.85$\,\mu_B$,
where $\mu_B$ is the Bohr magneton; nearly 85\% of this moment 
is derived from electrons in the uppermost three valence bands.
A closer examination indicates that each of the four O ions contribute 
as much as 0.09$\,\mu_B$/ion to the total magnetic moment of the CuO$_4$ unit. 
The development of such large magnetic moments on oxygen ions has been
previously predicted in Li$_2$CuO$_2$, a cuprate containing one-dimensional
CuO$_2$ chains.\cite{pickett} We also note that our calculation compares
well with the
most recent experimental value of $\sim$0.8$\,\mu_B$ extracted from
neutron Bragg-peak intensities of CCTO,\cite{ram3} although less favorably
with a previous value of $\sim$0.96$\,\mu_B$.\cite{collomb}
Both, however, are significantly larger (by nearly a factor
of 2) than moments seen in some undoped cuprates, such as
La$_2$CuO$_{4-y}$,\cite{vak} which are notably parent compounds
of the doped high-$T_c$ superconductors.

We calculate an indirect band gap of 0.19\,eV\cite{gap} and a minimum
direct gap of 0.27\,eV.  Our computed gap thus greatly underestimates the
experimental optical gap, which must exceed 1.5\,eV according to recent
data.\cite{pcom}
That the LSDA underestimates the gap dramatically in this case is not
particularly surprising; the localized $3d$ orbitals centered on the
Cu$^{2+}$ ions will almost certainly result in strong on-site Coulomb
interactions, which are not adequately treated at the level of the LSDA.
In fact, we expect these interactions to be sufficiently strong that the
material is best described as a Mott-Hubbard insulator.  In view of this, 
some additional details of our calculated electronic and magnetic
structures should not necessarily be taken at face value.  For example,
we computed the total energies of the ferromagnetic (FM) and unpolarized
paramagnetic (PM) states and found them to be higher than the energy of
the AFM ground state by 30 and 60\,meV/Cu ion, respectively.  
Further, both our FM and PM states are {\it metallic}.
Because of the inadequacies of the LSDA, 
the temptation to use the AFM--FM splitting to
estimate the Weiss constant or N\'eel temperature should be avoided; both would
be unphysically large. Moreover, within
the conventional picture of a Mott-Hubbard insulator we expect the
system to remain insulating, with little change in the optical gap, 
as the system is switched from an
AFM to a FM state, or even as the temperature is raised through the
N\'eel temperature to create a disordered spin state.

Clearly, the LSDA has serious limitations for a system of this
kind, and therefore it would be of value to explore the electronic
and magnetic structure of this system using more sophisticated
many-body tools.  One such direction would be along the lines
of quasiparticle theory, perhaps using the Bethe-Salpeter
extension \cite{bslouie} of the GW theory.\cite{gw}
An alternative approach could be along the lines of LSDA+U \cite{ldau}
or dynamical mean-field theory (DMFT)\cite{dmft}.  These approaches,
which have the capability of taking strong electron-electron interactions
into account, would presumably give a more realistic description of
the electronic ground state and its excitations in a material
such as CCTO.

\section{ZONE-CENTER OPTICAL PHONONS}

In most insulating perovskites such at SrTiO$_3$ and CaTiO$_3$,
the dominant contribution to the static dielectric constant comes
from the lattice susceptibility associated with the IR-active
phonon modes.\cite{Cockayne}  Thus, we are strongly motivated to
calculate the frequencies and dynamical mode effective charges of
these zone-center optical phonon modes, in order to test whether
the lattice susceptibility can explain the enormous observed static
dielectric constant of CCTO.  Moreover, the zone-center optical modes
are also of interest in view of some seemingly anomalous behavior
observed in the optical IR and Raman spectra as a function of
temperature.\cite{ram,homes}
Only ten IR-active modes are observed, despite the fact that
the symmetry of the crystal requires eleven distinct IR-active
optical phonons. And, perhaps more crucial, the lowest peak,
with a frequency of $\sim$122\,cm$^{-1}$ at room temperature,
shifts to slightly lower frequencies, broadens, and develops a
pronounced shoulder as the temperature decreases. Interestingly,
the oscillator strength associated with this peak also {\it
increases} by a factor of 3.6, in apparent violation of the $f-$sum
rule.   Raman measurements\cite{ram} reveal not only three sharp phonon
lines, but also some weaker features and a peculiar continuum
extending from about 210 to 420\,cm$^{-1}$.  It is unclear
whether or not these IR and Raman anomalies are connected with
the huge static dielectric response, but it is clearly of interest
to investigate them as far as possible.

With these motivations in view, we calculate the frequencies
of all zone-center optical phonon frequencies in CCTO, and for
IR-active modes we also compute the dynamical effective charges
needed to evaluate the lattice contribution to the static
dielectric constant.

\subsection{Symmetry analysis}

Since the Hamiltonian remains invariant under
a lattice translation of the 20-atom primitive bcc cell with an
accompanying spin reversal, the force constant matrix elements,
which are the second derivatives of the total energy with
respect to atomic displacements only, are also unchanged. Thus
the force-constant matrix has the translational symmetry of the
20-atom cell, and in what follows, we use the symmetry-adapted
modes determined from the 20-atom primitive cell (and not the
40-atom conventional cell) in our analysis of zone-center phonons.

To calculate the phonon contribution to the dielectric response,
we need only determine frequencies and mode effective charge
of the IR-active phonons at the Brillouin-Zone (BZ) center.
Altogether there are 60 distinct modes at the $\Gamma$ point,
including the three acoustic (zero-frequency) modes.  To minimize
the numerical burden, we first decompose these modes into the 6
allowed irreducible representations:
\begin{equation}
\Gamma = 2\mbox{A}_g \oplus 2\mbox{E}_g \oplus
 4\mbox{T}_g \oplus 2 \mbox{A}_u \oplus 2 \mbox{E}_u \oplus 12 \mbox{T}_u
\end{equation}

Symmetry-adapted modes transforming as each of the irreducible
representations given above were constructed using the SMODES
symmetry-analysis software package.\cite{iso} Among those phonon
modes, only the $\mbox{T}_u$ modes are infrared (IR) active;
the $\mbox{A}_u$ and $\mbox{E}_u$ are silent, and the remaining
modes are Raman active.

\subsection{Zone-center phonon frequencies}
\label{sec:phonons}

For each irreducible representation $\Gamma$, we construct an
$N_{\Gamma}\times N_{\Gamma}$ dynamical matrix from a series of
frozen-phonon calculations in which the structure is distorted
according to each of the $N_{\Gamma}$ symmetry-adapted coordinates
consistent with this irreducible representation.  For example,
for the case of the $N_{\Gamma}$=12 IR-active T$_u$ modes (each
of which is three-fold degenerate), we set up a 12$\times$12
dynamical matrix.  In all cases, the forces are calculated on
the ions in the 40-atom cell and with respect to the correct AFM
reference ground state within the LSDA. After calculating the
residual Hellmann-Feynman forces $F_i^{\alpha}$ on ion $i$ and in
Cartesian direction $\alpha$ due to displacements $u_j^{\beta}$
of ion $j$ in direction $\beta$, we displace the ions according to
the symmetry coordinate for each mode by 0.1\% of lattice constant
and recalculate the forces. The force constant matrix
\begin{equation}
\Phi^{\alpha\beta}_{ij} = -{\partial F_i^{\alpha} \over \partial u_j^{\beta}} 
\end{equation}
is then obtained by finite differences by freezing in a small
amplitude of each symmetry-adapted distortion pattern that contributes
to the modes of a given symmetry and computing the resulting forces
that arise.  The normal modes ${\bf u}_\lambda$ and their frequencies
$\omega_{\lambda}$ are then obtained through solution of the
eigenvalue equation
\begin{equation}
{\bf \Phi}\cdot{\bf u}_\lambda =\omega_\lambda^2 \,
{\bf M}\cdot{\bf u}_\lambda.
\end{equation}
Here $M_{ij}^{\alpha\beta}=(m_i/m_0)\,\delta_{ij}\,\delta_{\alpha\beta}$
is the dimensionless diagonal mass matrix, where $m_i$ is the mass
of atom $i$ and $m_0$ is a reference mass chosen here to be
1\,amu,\cite{explan-mass} and the eigenvectors are normalized
according to ${\bf u}_\mu \cdot {\bf M} \cdot {\bf u}_\nu =
\delta_{\mu\nu}$.

The results for the IR-active modes are shown in Table
\ref{tab:IR}. The overall agreement with experiment\cite{homes} 
is excellent, except for one mode ($\omega=$483\,cm$^{-1}$) apparently missing
in the measurement.  By symmetry, there should be 11 optical modes
(plus 1 acoustic mode) in total, yet only 10 optical modes were observed by experiment.
It is possible that the oscillator strength for this mode is immeasurably small;
alternatively, its frequency may be
so close to that of another mode ($\omega=$507\,cm$^{-1}$)
that the pair of modes appears together as a single peak in
the optical data.  As we show in the next section, however, we
compute an appreciable, non-zero oscillator strength for this mode,
which is mainly composed of displacements of Ti and O ions.  Furthermore,
the sum of the calculated oscillator strengths of the modes at both
483\,cm$^{-1}$ and 507\,cm$^{-1}$ would greatly exceed that obtained
from the observed peak.  We return to these inconsistencies below
in Sec.~VI\,D.

\begin{table}
\caption{Calculated mode frequencies $\omega_\lambda$,
effective charges $Z_{\lambda}^*$, and oscillator strengths
$S_{\lambda}=\Omega_0^2\,Z_{\lambda}^{*2}/\omega_{\lambda}^2$
of IR-active T$_u$ modes, compared with experimental values extracted from 
Refs.~\protect\onlinecite{homes}.}
\label{tab:IR}
\vskip 0.1cm
\begin{tabular}{dddddd}
\multicolumn{2}{c}{$\omega$ (cm$^{-1}$)} &
\multicolumn{2}{c}{$Z^*_\lambda$} &
\multicolumn{2}{c}{$S_\lambda$} \\
LSDA & \multicolumn{1}{c}{Exper.} &
LSDA & \multicolumn{1}{c}{Exper.} &
LSDA & \multicolumn{1}{c}{Exper.} \\
\hline
125	& 122.3    & 0.51  & 0.94   &  4.1   & 14.3 \\
141	& 140.8    & 0.65  & 1.12   &  5.4   & 15.9 \\  
160     & 160.8    & 0.91  & 0.84   &  8.1   & 6.92 \\
205     & 198.9    & 0.55  & 0.93   &  1.8   & 5.25 \\
264     & 253.9    & 1.79  & 1.95   &  11.5  & 13.8 \\
314     & 307.6    & 0.66  & 0.52   &  1.1   & 0.68 \\ 
394     & 382.1    & 0.36  & 1.1    &  0.2   & 1.96 \\
429     & 421.0    & 1.83  & 1.12   &  4.6   & 1.72 \\
483     &          & 1.44  &        &  2.2   &      \\
507     & 504.2    & 0.67  & 0.89   &  0.4   & 0.78 \\
563     & 552.4    & 0.73  & 0.88   &  0.4   & 0.62
\end{tabular}
\end{table}

We also calculate the frequencies of the Raman-active modes
and silent modes; the results are listed in Table
\ref{tab:Raman}.  In the case of the Raman modes, we also give
proposed assignments of the experimentally observed modes.
In the Raman experiment\cite{ram}
the incident and scattered light were polarized parallel to each
other,\cite{ramirezprivate} in which case the symmetry of the Raman
tensor requires the intensity of all T$_g$ modes to vanish.\cite{loudon}
The experimental spectrum\cite{ram} shows three clear peaks at 445,
513, and 577\,cm$^{-1}$.  In addition, there is an anomalous
temperature-dependent continuum extending from about 210 to
420\,cm$^{-1}$, on which is superimposed a weak peak at approximately
290\,cm$^{-1}$.  Finally, there is also a weak broad peak at
approximately 760\,cm$^{-1}$.  The three strong peaks are easily
assigned as in Table \ref{tab:Raman}, and the weak low-frequency
peak is almost certainly associated with the E$_g$ mode
calculated to lie at 292\,cm$^{-1}$.  The feature observed
at 760\,cm$^{-1}$ could be attributed to the T$_g$ mode computed at
739\,cm$^{-1}$ if the assumed polarization alignment was imperfect,
allowing some leakage of T$_g$ mode intensity into the spectrum.
We cannot directly explain the continuum observed in the
210-420\,cm$^{-1}$ range, but note that it could result from a
two-phonon Raman process, possibly involving pairs of IR-active
modes.  Overall, we find excellent agreement between theory and
experiment for the frequencies of Raman-active modes.

To summarize, we have computed the frequency of all zone-center
modes of CCTO and confirmed that all are stable ($\omega^2>0$),
consistent with experiments and with our structural relaxation.
In particular, our computed ground state is confirmed to be locally
stable against all possible ferroelectric lattice distortions.

\begin{table}
\caption{Frequencies (in cm$^{-1}$) of Raman-active and silent modes
of CCTO, with proposed assignments for the three modes observed
experimentally in Ref.~\protect\onlinecite{ram}.  Irreducible
representations are given in parentheses.}
\label{tab:Raman}
\vskip 0.1cm
\begin{tabular}{lcc}
     & LSDA &Exper.\\
\hline
Raman modes
       & 277 (T$_g$)   &     \\
       & 292 (E$_g$)   &     \\
       & 437 (T$_g$)   &     \\ 
       & 439 (A$_g$)   & 445 \\
       & 519 (A$_g$)   & 513 \\
       & 552 (T$_g$)   &     \\
       & 568 (E$_g$)   & 577 \\
       & 739 (T$_g$)   &     \\
\hline
Silent modes
       & 123 (E$_u$) &    \\
       & 179 (A$_u$) &    \\
       & 472 (A$_u$) &    \\ 
       & 490 (E$_u$) &    
\end{tabular}
\end{table}

To test whether the $Im3$ structure is truly a global or only a local minimum,
we also calculated the forces and frequencies along the direction
defined by some phonon modes at different displacement amplitudes,
including the lowest frequency IR phonon. It is found that these
phonons remain harmonic for amplitudes up to 4$\%$ of the
lattice constant. We also examined the possibility of coupling
between the low-frequency 122\,cm$^{-1}$ mode (${\bf u}_{\rm IR}$)
and the Raman modes ${\bf u}_{\rm Raman}$ ($\omega=$277cm$^{-1}$
and 292\,cm$^{-1}$), which have frequencies roughly twice that of
the lowest-frequency IR mode, to see if multi-phonon processes
(i.e., anharmonicity) could explain the anomalies present in the
Raman spectrum and optical response measurements.  For each of the
two lowest-frequency Raman modes, we calculate the total energy
versus displacement along two separate directions defined by
both ${\bf u}_{\rm IR} + {\bf u}_{\rm Raman}$ and
${\bf u}_{\rm IR} - {\bf u}_{\rm Raman}$ combinations
of these two modes. Our calculations along these four directions
(two associated with each Raman mode) indicate that the total
energy is quite quadratic for a wide range of amplitudes (up to
4\,\% of the lattice constant), indicating that the coupling is too
weak among these modes to explain the magnitude of the observed
anomalies. Our calculations do not, however, absolutely rule
out the possibility of significant coupling between more general
combinations of phonons. A more complete stability analysis would
require the calculation of the full phonon spectrum; we leave
this formidable task for a future investigation.

\subsection{Effective charges and lattice dielectric response}

We now turn our attention to a calculation of the lattice contribution
to the static dielectric constant, in order to test whether the
predicted response is of the correct order of magnitude to explain
the enormous observed susceptibility.  For this purpose, and also
motivated to understand better the observed anomalous temperature
dependence of the oscillator strength of the 122\,cm$^{-1}$ mode
(see Sec.~\ref{sec:anomalies}), we calculate the effective charges
of the IR-active modes and their contribution to the dielectric
constant.  The dielectric function in the frequency range near the
lattice vibrations can be written as a sum
\begin{equation}
\epsilon(\omega) = {\epsilon_{\infty}} + \epsilon_{\rm ph}(\omega)
\end{equation}
of electronic and lattice contributions.  In most
insulating cubic perovskite oxides (e.g., ferroelectrics
and related materials) one has $\epsilon_\infty\sim 5$ and
$\epsilon_0$=$\epsilon(0)\sim$20-100,
so that the lattice contribution is expected to dominate.
It can be written
\begin{equation}
\epsilon_{\rm ph}(\omega) = {\Omega_0}^2 \sum_\lambda 
{{{Z^*_\lambda}^2}\over{{\omega_\lambda}^2 - \omega^2}} \;,
\end{equation}
where  $\omega_\lambda$ and $Z^*_{\lambda}$ are respectively the
mode frequencies and mode dynamical charges, and ${\Omega_0}^2= 4\pi
e^2/m_0 V$ is a characteristic frequency having the interpretation
of a plasma frequency of a gas of objects of mass
$m_0$=1\,amu,\cite{explan-mass} charge $e$, and density $V^{-1}$
($V$ is the 20-atom primitive cell volume).
Here, we have used the cubic symmetry of CCTO to restrict the
sum over $\lambda$ to just one member of each of the 11 three-fold
degenerate sets of IR-active T$_u$ modes, namely the one polarized
along $z$, and the mode dynamical charge is then expressed in
terms of atomic dynamical effective charges $Z_{i\alpha\beta}^*$ as
\begin{equation}
Z^*_\lambda = \sum_{i\alpha} Z_{i\alpha z}^* \, u_{\lambda i}^{\alpha}
\;.
\label{eq:zlambda}
\end{equation}

In practice we compute the mode dynamical charge by finite
differences as $Z^*_\lambda = {V \Delta P_z /u_0}$, where
$\Delta P_z = P_z({\bf u}_{\rm eq} + u_0 \, {\bf u}_{\lambda})-
P_z({\bf u}_{\rm eq})$ and $u_0$ is a small frozen-ion amplitude
of the normal mode ${\bf u}_{\lambda}$ added to the equilibrium
position of each ion.
$\Delta P$ has both ionic and electronic contributions.  The ionic
contribution is obtained by summing the product of the displacement
of each ion with the nominal charge of its rigid core; the
electronic contribution is computed from first principles using the
Berry-phase theory of polarization.\cite{ksv}  More specifically,
the electronic part of the polarization is determined to within
a constant (an integer multiple of the polarization ``quantum''
$e{\bf R}/V$, where ${\bf R}$ is a lattice vector) by evaluating
the phase of the product of overlaps between cell-periodic Bloch
functions at successive {\bf k}-points along a densely-sampled
string of points in {\bf k}-space.  This string is chosen
so it is parallel to the direction of the desired polarization.
For each calculation we use one symmetrized string consisting
of 4 {\bf k}-points (corresponding to 8 {\bf k}-points/string
in a 5-atom perovskite cell).  Increasing the number of {\bf
k}-points along the string from 4 to 8 changes the effective
charges by less than 1\%.  The electronic contribution to the
polarization is calculated separately for each spin channel; the
total polarization is then the sum of the two spin contributions
with the ionic contribution.

The calculated mode effective charges are listed in Table II, where
they are compared to values obtained from the optical conductivity
measurement of Homes {\it et al}.\cite{homes} 
Surprisingly, though the agreement is excellent between
our calculated {\it frequencies} and those measured at
room temperature, the theoretical values of the {\it mode
effective charges} are in poorer agreement with the measured
oscillator strengths. That the experimental values refer to room
temperature and the computations to 0\,K may account for some of
the discrepancies, although the large increase measured in the
oscillator strength of the lowest-frequency mode at low temperature
would compare even less favorably with our calculations. In
this context, we note that the measured oscillator strengths are
strongly temperature-dependent, while the mode frequencies are
only weakly so, suggesting a strong temperature-dependence of
the mode effective charges.  Because the mode effective charges
are essentially electronic quantities, these observations could
possibly suggest that the electronic structure of this compound is unusual
in some way, and perhaps that this peculiarity is related to the
enormous dielectric response.  These issues will be discussed
further in the next section.

Having these mode effective charges, we estimate the lattice
contribution to the static ($\omega\rightarrow 0$) dielectric constant
$\epsilon_{\rm ph}(0) = \Omega_0^2  \sum_\lambda Z_{\lambda}^{*2}
/ \omega_{\lambda}^2 =\sum_\lambda S_\lambda$.
Using the eigenmodes and their effective
charges in Table II, we obtain $\epsilon_{\rm ph}$$\sim$40
(at zero temperature).  Assuming a typical $\epsilon_\infty$$\sim$5-10,
we expect $\epsilon_0$$\sim$45-50, to be
compared with a sub-IR-frequency dielectric constant of
$\sim$70 at room temperature and $\sim$120 at 10\,K.
Thus despite some detailed disagreement between
the experiment and theory for the mode effective charges, the
total contribution of these modes to the dielectric constant is
of roughly the right order of magnitude in this frequency range.

\section{DISCUSSION AND COMPARISON WITH EXPERIMENT}

Although we obtain satisfactory agreement with experiment 
for computed structural, electronic, magnetic, and lattice-dynamical properties, 
our calculations {\it do not} suggest any obvious
explanation for either the low-frequency Debye relaxation or its
associated enormous static dielectric response. 
In addition, the observed anomalies associated
with the temperature-dependence of the IR and Raman phonons also remain
unexplained.  The strong $T$-dependence of the oscillator strength
of the lowest-frequency IR-active mode is especially strange.
In this section we consider various possible explanations for these
distinctive features of CCTO and suggest further experiments
that might shed new light on these puzzling phenomena.  

\subsection{Possible origins of the large dielectric response and
low-frequency Debye relaxation }

We consider two
classes of possible explanations for the dielectric response 
and its frequency dependence, one primarily {\it intrinsic} 
and one primarily {\it extrinsic}.
By intrinsic, we mean the response that would be measured in a
perfectly stoichiometric, defect-free, single-domain crystal of CCTO.
Extrinsic effects, on the other hand, would be those associated with defects 
or other crystal imperfections.

\subsubsection{Intrinsic mechanisms}

We first consider the possibility that the origin of the dielectric
response is actually intrinsic, in which case the fact that our
calculations do not reproduce the observed behavior could be ascribed
to some approximation or restriction of our calculation (e.g., the
local spin-density approximation or the fact that we work at zero
temperature).

One class of intrinsic origins for the anomolous response
would be related to a lattice instability 
resulting in ferroelectric or relaxor-like 
behavior. The anomalies recently observed in both the IR and Raman spectra
indeed raise the possibility of a close connection between the lattice modes
and the enormous dielectric response. 
Yet, in view of existing experimental evidence and the results of 
our calculations discussed above, 
any such connection between the large dielectric
response and lattice dynamics cannot be described by simple ferroelectricity. 
Additionally, 
there are several specific objections to a relaxor picture, even from
a purely experimental point of view.  
First, relaxors are typically solid solutions (e.g., PMN =
Pb(Mg$_{1/3}$Nb$_{2/3})$O$_3$) in which the random identity of atoms
on one of the sublattices provides an obvious source for disorder.
In CCTO the Ca/Cu sublattice is well ordered, and hence there is no
obvious source for relaxor-like disorder.  Second, it is very hard
to reconcile a picture of this kind with the diffraction
experiments, all of which fail to observe any of the broadening,
superstructure, reduced Bragg intensities, or diffuse background
that should be associated with such structural disorder.  Third,
either the domains are small enough to fluctuate freely in orientation,
or they are large enough so that reorientation of their polarization
can occur only by domain-wall motion. In the former case, 
a temperature independent dielectric constant is unlikely. 
In the latter case, the magnitude of the dielectric constant should
be comparable only to that observed in other ferroelectrics containing domains, 
unless the local polarization is unusually large or 
the domain-wall mobility monumentally facile.

Moreover, a relaxor picture is difficult to reconcile
with our theoretical calculations, since it requires the
structure to be unstable to some kind of local polar lattice distortion,
and we do not observe such instabilities in our calculations.  Of course, one
could argue that one of the IR modes really is soft ($\omega^2<0$) even though the
LSDA calculation indicates otherwise ($\omega^2>0$).  However, our
computed phonon frequencies for the undistorted $Im3$ structure
are in excellent agreement with the experimentally observed ones.

Another possible intrinsic origin of the observed dielectric response
could be a novel, highly-correlated electronic ground state.
Given that CCTO is an antiferromagnetic Mott insulator with an electronic
structure containing relatively localized $d$ states as well as dispersive $s$ and $p$
bands, one possibility would be that its anomalous dielectric response
could be explained by the formation of {\it electronic
ferroelectricity} (EFE).\cite{fk,sham,czycholl} As it is unlikely
that the LSDA calculations would correctly describe such a state,
our prediction of a conventional antiferromagnetic ground
state should not be taken as definitive evidence against this
scenario.  Nonetheless it can be shown based on symmetry considerations
that a {\it purely} electronic ferroelectric state
{\it is not possible}.  In an EFE state, quantities such as the
electronic charge density and one-particle density matrix would
break the $T_h$ point-group symmetry (and in particular, inversion
symmetry).  Since the IR-active optical phonons would necessarily couple to such a symmetry
breaking, we expect these modes to develop nonzero static
mode amplitudes in the EFE state, a result that would be evident
in the diffraction or optical spectra (unless the coupling is
unusally small).

Several more specific objections to other possible electronic origins 
can also be raised.  First, an electronic instability is 
most likely in a system with a small gap.  However, 
the experiments suggest that the optical gap in CCTO
exceeds 1.5\,eV, as would be consistent with the picture of a
conventional Mott insulator.  It is hard to understand how excitons or
other low-frequency excitations could condense in a system with
such a robust gap.  Second, regarding our theoretical calculations,
it is difficult to reconcile our reproduction of the observed phonon frequencies 
with the inability of the LSDA to reproduce the qualitative nature of the
electronic ground state.  Third, the Debye relaxation time $\tau$, which is
observed to be activated with an attempt frequency in the MHz to
GHz range, appears to be too slow to attribute to an electronic mechanism.

\subsubsection{Extrinsic mechanisms}

In view of the experimental and 
theoretical difficulties with an intrinsic explanation,
we observe that many puzzling aspects of the behavior of CCTO become easier
to understand if one assumes that the large observed
static dielectric response arises from extrinsic mechanisms, such
as point, line, or planar defects, or more generally, with sample
microstructure, morphology, or boundary layers.
In this case, the intrinsic static dielectric
susceptibility would actually be on the order of $\sim$70--120, as is typical
for conventional dielectrics in the same structural class.\cite{sub}  
The large response would then be consistent with our theoretical calculations, and
with most of the experimental observations (x-ray and neutron
diffraction, IR and Raman phonon spectra, etc.) that would effectively observe
only the conventional dielectric material.  In this scenario, the
observed anomalies in the IR and Raman spectra and
the huge Debye dielectric response would be completely unrelated phenomena.

Since the dielectric response seems to be too large to be explained by point (or line)
defects alone, boundary or interface effects would be needed.  
There are two obvious candidates for such internal 
interfaces even in ``single-crystal'' samples:  
antiphase boundaries separating regions in which the
identity of the Ca sublattice is shifted by a primitive-cubic
lattice vector, and twin boundaries separating domains in which the
sense of rotation of the oxygen octahedra reverses. 
Diffraction data\cite{ram} does not rule
out the occurrence of either kind of domain wall.

If such domain boundaries exist, they could possibly give rise to a
large dielectric response as follows.  Suppose that, away from such
boundaries, CCTO actually has a weak bulk conductivity $\sigma_{\rm
b}$ resulting from point defects of some kind.  Moreover, assume that the
conductance of the sample as a whole is eliminated by the presence
of insulating antiphase or twin boundaries that divide the sample
into non-percolating, conducting domains.  A static dielectric
constant on the order of $\epsilon_{\rm ins}/f$ would then be
expected, where $\epsilon_{\rm ins}$ is the dielectric constant of
the insulating interlayer phase and $f$ is its volume fraction.  Assuming
$\epsilon_{\rm ins}\sim10^2$ and $f\sim10^{-3}$, an overall static
dielectric constant of $10^5$ is quite plausible.  Moreover,  the
observed {\it temperature-independence} of this macroscopic response
is easily explained as long as $\epsilon_{\rm ins}$ and $f$ do not
vary significantly with temperature (in particular, 
a strongly temperature-dependent $\sigma_{\rm b}$ is allowed).
Such a model predicts a Debye relaxation
behavior, with a relaxation time $\tau$ scaling as $\sigma_{\rm
b}^{-1}$, and the observed activated behavior of $\tau^{-1}$ is
then naturally ascribed to an activated behavior of the intradomain
conductivity $\sigma_{\rm b}$.\cite{morrel}  The observed activation energy of
54\,meV could be consistent with a variety of mechanisms for such
an activated conductivity. For example, one could assume electronic conductivity
associated with gap states arising from stoichiometric variations
or dopant impurities, or even ionic conductivity associated with
mobility of charged defects.  This model, which is essentially an
elaboration of a suggestion made by Subramanian {\it et al.} in a
``Note added in proof'' in their Ref. 1, is worked out in further detail
elsewhere\cite{morrel} for various morphologies of the blocking
boundaries, including the possibility that the blocking occurs at the contacts
instead of at internal boundaries.

Perhaps the trickiest aspect of such models is that they require the
insulating blocking layers to be so uniform and robust that they
completely prevent conductivity of the sample as a whole, despite
their small volume fraction.  One would then expect that small variations
in sample preparation procedures might generate samples having enough
``holes'' in the blocking layers that a measurable conductivity
could be observed.  In fact, there are hints of such behavior in the
experimental work.\cite{ramirezprivate} 
Further, if the enormous dielectric constant of CCTO does in fact originate from
the suggested extrinsic mechanism, the width of the blocking layers,
and thence the size of the dielectric response, might be controllable.
Accordingly, it would be interesting to explore the internal morphology
of single-crystal CCTO samples, both theoretically\cite{morrel} and
experimentally.  In this context, we also note that recent studies have
uncovered two other new materials, Ca-doped KTa$_{1-x}$Nb$_x$O$_3$ (KTN:Ca)\cite{samara} and
La-doped PbTiO$_3$ (PLT-A)\cite{kim}, which also exhibit Debye-like relaxation.
However, in KTN:Ca this relaxation occurs at temperatures {\it
above} the ferroelectric transition; in PLT-A it occurs at
temperatures {\it below} the ferroelectric transition; and in CCTO
it evidently appears in the {\it absence} of a ferroelectric
transition.  This would further indicate that the relaxation process
is not necessarily related to ferroelectric or relaxor-like behavior 
at all. 

\subsubsection{Suggestions for experimental tests}

Several types of experiments might be useful in
testing whether any of the mechanisms proposed above are at work in
CCTO.  First, it would be interesting to test whether, when looking
at sample-dependent variations, one observes any correlation
between the behavior of the huge dielectric relaxation (e.g.,
$\epsilon_0$ values) and properties that are
clearly intrinsic (e.g., the oscillator strength of the
lowest-frequency IR mode).  Such a correlation would
tend to rule out an extrinsic mechanism and establish a connection
between the phonon anomalies and the huge dielectric response.
Second, for samples prepared in such a way as to have a measurable
DC conductivity, it would be very interesting to test whether
their conductivity is activated in the same way as required to
explain the Debye behavior in the macroscopically nonconducting samples.
Third, C--V measurements might give strong indications about which
type of model is correct.  Models based on lattice or
electronic ferroelectricity would tend to
predict a saturation of the polarization at a characteristic bulk
``spontaneous polarization,'' while the extrinsic models predict a
more-or-less linear behavior up to some kind of breakdown voltage.
Nonlinearities would reveal the importance of anharmonic effects.

Some other possible avenues of future experimental investigation
would be to use electronic probes such as angle-resolved
photoemission and inverse photoemission to study the nature of the
valence and conduction band edges and to look for unusual
quasipartical behavior, or to use optical measurements to characterize
the dielectric response in the vicinity of the optical edge.
Experiments that could be sensitive to local structural distortions (e.g., EXAFS,
nuclear electric quadrupole, etc.) or to glassy
dynamics (NMR, spin echo) would also be worth pursuing.  An isotope
experiment may provide essential evidence for or against a lattice mechanism.
Experiments designed to image the assumed
domain-boundary blocking layers (e.g., TEM), and to determine their chemical and
physical properties, would obviously be of interest.

\subsection{Experimental IR anomalies}
\label{sec:anomalies}

We return here to a discussion of the anomalous behavior of the
low-frequency peak in the IR response observed by Homes et
al.\cite{homes}  This mode, which appears at 122\,cm$^{-1}$ at room
temperature, is observed to broaden and shift to lower frequency
($\sim$115\,cm$^{-1}$ at 10\,K) as the temperature is reduced, where
it appears to develop a high-frequency shoulder. Most
surprisingly, however, its oscillator strength is found to
increase by almost a factor of four as the temperature drops from 300\,K
to 10\,K.  Moreover, this increase does not seem to be offset by a
decrease of oscillator strength of other modes; in fact, the total
oscillator strength summed over all IR modes is also observed to rise,
growing by $\sim$10\% from 300\,K to 10\,K.  This seems
especially puzzling since it appears to result in a 
violation of an oscillator-strength $f$-sum rule.

To be precise, it is possible to show that, provided there is a good
adiabatic separation between the frequency scales of lattice and
electronic responses, one expects a sum rule of the form
\begin{equation}
{2\over\pi} \int {\rm Im}\,\epsilon(\omega)\,\omega\,d\omega
= \Omega_{\rm p}^2
\label{eq:sumrule}
\end{equation}
to hold, where the integral runs over the frequency domain of the lattice
modes and
\begin{equation}
\Omega_{\rm p}^2=\sum_i \, {4\pi e^2 \over V} \,
{\langle Z^{*2}_i\rangle \over M_i}
\label{eq:plasmafreq}
\end{equation}
has the interpretation of a ``plasma frequency'' associated with
the ionic charges and masses.  Here $i$ runs over atoms in the unit
cell of volume $V$, $Z_i^*$ is the Born effective charge 
(isotropically averaged \cite{explan-zstar}) of atom $i$, and the
angle brackets indicate a thermal ensemble average.  This result
follows directly from the Kramers-Kronig relations and the
high-frequency lattice dielectric response ${\rm Re}\,\epsilon(\omega)
\simeq \epsilon_\infty-\Omega_{\rm p}^2/\omega^2$ expected
in the frequency region above the lattice modes, where the equations
of motion are dominated by electric forces and inertia.

The observed strong temperature-dependence of the left-hand side of
Eq.~(\ref{eq:sumrule}) need not violate any physical law.  Rather,
it suggests one of two things.  First, it may simply be that the
atomic $Z^*$ values are strongly dependent on the local
displacements of the atoms, in such a way that the thermal
expectation value on the right side of Eq.~(\ref{eq:sumrule})
acquires a strong temperature-dependence.  Second,  
if some kind of continuum of low-frequency electronic
excitations extends down to the frequency range of the lattice
modes, then the assumptions underlying the derivation of
Eq.~(\ref{eq:sumrule}) are not valid.  In this case, the electronic
structure (at fixed ionic coordinates) might be strongly
temperature-dependent in the temperature range of the anomaly.

Our theoretical calculations do not provide any guidance in
distinguishing between these possibilities.  The second possibility
could be suggestive of some kind of electronic mechanism for the
enormous dielectric response.  
However, we note that the temperature and frequency ranges of the
IR anomalies are rather different from those of the Debye relaxation,
so that such a connection is not necessarily expected.  Perhaps
careful and precise measurements of $\epsilon(\omega)$
in the frequency range above all phonon modes (i.e., above
750\,cm$^{-1}$) would be the best way of testing whether or not the
assumed adiabatic separation between lattice and electronic degrees
of freedom really does occur. Additionally, measuring the oscillator
strengths as a function of {\it pressure} may aid in uncovering the nature of
the missing mode, as well as providing further information about
the anomalous temperature-dependence of the oscillator strengths
of the other modes.

\section{CONCLUSIONS}

We have performed first-principles calculations within the
local-density approximation on CaCu$_3$Ti$_4$O$_{12}$ in order to gain
insight into the enormous dielectric response reported in recent
experiments.  Our calculated ground-state structural properties
and phonon frequencies agree quite well with preexisting
measurements.  The electronic structure we calculate appears to
underestimate the empirical optical gap in a manner consistent
with previous calculations within the LSDA. Our computed effective
charges are of the correct order of magnitude or better and result in a
lattice contribution to the dielectric constant that is roughly
consistent with that observed in the IR range.

While this work does not provide a theoretical explanation for either
the Debye relaxation or the unusually large and temperature-independent
dielectric response that is observed,
our calculations do appear to limit certain intrinsic
mechanisms and point toward an extrinsic cause.
We find CCTO to be stable in a centrosymmetric crystal structure
with space group $Im3$, arguing against the possibility that CCTO is
a conventional ferroelectric or relaxor.  Our use of the local-density
approximation constrains our ability to rule out more
unconventional, purely electronic mechanisms, though we note
that there is currently no evidence of electronic excitations in
the optical spectrum. We also add that there is no experimental
support from existing structural or electrical measurements for the
large dipole moment, whether electronic or displacive in origin,
required in each unit cell to produce the anomalous response.
We are unable to provide a quantitative explanation for the
changes in oscillator strength of the low-frequency IR-active mode
observed in experiment as a function of temperature, although,
on the basis of existing measurements, we speculate that it
may be uncorrelated with the large dielectric response.

An extrinsic mechanism for the static dielectric response and Debye
relaxation has many attractive features.  In one version of such
a mechanism, a weak conductivity in the interior of domains is barely
prevented from leading to a macroscopic bulk conductivity by the
presence of then interface blocking layers.\cite{morrel}  In such a
scenario, the dielectric susceptibility at and below the Debye response
frequency is controlled by this internal conductivity, and is unrelated
to lattice distortions, structural phase transitions, or novel
electronic states of the crystal.  Such a mechanism is then consistent
with the absence of diffraction evidence for the kinds of
symmetry-lowering distortions expected if a lattice mechanism were at
work.  It is also consistent with the observation that our computed
zone-center phonon frequencies are in excellent agreement with
experiment.  By the same token, however, such a mechanism cannot
explain the observed phonon anomalies, which would have to be
attributed to an unrelated cause.

In any case, it is of the utmost importance to clarify which type of
mechanism really is responsible for the huge dielectric response and the
Debye relaxation behavior in CCTO.  For this purpose, the ultimate test
must be experimental.  We have suggested numerous future
avenues of experimental as well as theoretical investigations that
could help settle this question.  It is our hope that the work
described here can serve as a firm theoretical foundation for
future investigations of this intriguing material.

\acknowledgments

This work supported by NSF Grant DMR-9981193.
We would like to thank K. M. Rabe for valuable discussions,
and also A. P. Ramirez, and S. M. Shapiro for
communicating results prior to publication and for useful
discussions.  We acknowledge
M. Marsman for his implementation of the Berry phase technique
within VASP. This work is supported, in part, by the U.S. Department of
Energy, Division of Materials Science, under Contract 
No. DE-AC02-98CH10886.




\begin{references}
\bibitem{gap} The indirect gap is between the high-symmetry
points $R$ (the valence band maximum) and $X$ (the conduction
band minimum), using the standard notation of the cubic Brillouin
zone. The smallest direct gap is at the $X$ point.
\bibitem{explan-mass} Alternatively, $m_0$ can be removed from
all formulas, in which case $Z_\lambda^*$ and $\Omega_0$ acquire
units of amu$^{-1/2}$ and amu$^{1/2}$\,cm$^{-1}$ respectively.
\bibitem{explan-zstar} More precisely, $Z^{*2}_i$
in Eq.~(\ref{eq:plasmafreq}) should be interpreted as
${1\over3}\sum_\alpha Z^{*2}_{i\alpha}$, where $\alpha$
runs over Cartesian coordinates, and we have made use
of the $T_h$ point-group symmetry which insures that
$\epsilon_{\alpha\beta}(\omega)=\delta_{\alpha\beta}\,\epsilon(\omega)$.
\end{references}
\end{document}